\documentclass[11pt]{article}
\usepackage{amsmath,amsthm,amsfonts}
\DeclareMathOperator{\1}{id}
\DeclareMathOperator{\ii}{i}
\newcommand{\p}{\partial}
\renewcommand{\=}{\doteq}

\newcommand{\al}{\alpha}

\newcommand{\de}{\delta}
\newcommand{\x}{\mathbf x}
\newcommand{\y}{\mathbf y}

\renewcommand{\L}{\mathfrak{L}}
\newtheorem{thm}{Theorem}%[section]
\newtheorem{GLC}[thm]{Generalized Lie-Cartan Theorem}

\theoremstyle{definition}

\theoremstyle{definition}
 
\title{\bf Moufang transformations and \\Noether currents}
\author{Eugen Paal\\ \\
Department of Mathematics\\
Tallinn University of Technology\\
Ehitajate tee 5, 19086 Tallinn, Estonia\\ 
E-mail: eugen.paal@ttu.ee}
\date{}

\begin{document}
\maketitle
\thispagestyle{empty}

\begin{abstract}
The Noether currents generated by continuous Moufang tranformations are constructed 
and their equal-time commutators are found.
The corresponding charge algebra turns out to be a birepresentation of the tangent Mal'ltsev algebra 
of an analytic Moufang loop.
\end{abstract}

\section{Introduction}
The Noether currents generated by the Lie transformation groups are well known and widely exploited
in modern field theory and theory of elementary particles. Nevertheless, it may  happen that 
group theoretical formalism of physics is too rigid and one has to extend it beyond the Lie groups and algebras.  
From this point of view it is interesting
to elaborate an extension of the group theoretical methods based on
Moufang loops as a minimal nonassociative generalization of the group concept.
In particular, the Mal'tsev algebra structure of the quantum chiral gauge theory was
established in \cite{Jo,NS}.

In this paper, the Noether currents generated by continuous Moufang tranformations (Sec. 2)
are constructed.
The method is based on the generalized Lie-Cartan theorem (Sec. 3).
It turns out that the resulting charge algebra is a birepresentation of the tangent Mal'ltsev algebra of an
analytic Moufang loop (Sec. 4).  Throughout the paper $\ii\=\sqrt{-1}$.

\section{Moufang loops and Mal'tsev algebras}

A Moufang loop \cite{RM,HP} is a quasigroup $G$ with the unit
element $e\in G$ and the Moufang identity
\[
(ag)(ha)=a(gh)a,\qquad a,g,h\in G.
\]
Here the multiplication is denoted by juxtaposition.
In general, the multiplication need not be associative: $gh\cdot a\neq g\cdot ha$.
Inverse element $g^{-1}$ of $g$ is defined by
\[
gg^{-1}=g^{-1}g=e.
\]

A Moufang loop $G$ is said \cite{Mal} to be \emph{analytic} if $G$ is also a real
analytic manifold and main operations - multiplication and inversion map
$g\mapsto g^{-1}$ - are analytic mappings.

As in the case of the Lie groups, structure constants $c^{i}_{jk}$
of an analytic Moufang loop are defined by
\[
c^{i}_{jk}\=\frac{\p^{2}(ghg^{-1}h^{-1})^{i}}{\p g^{j}\p h^{k}}\Big|_{g=h=e}=-c^{i}_{kj},
\qquad i,j,k=1,\dots,r\=\dim G.
\]
Let $T_e(G)$ be the tangent space of $G$ at the unit element $e\in G$.
For any $x,y\in T_e(G)$, their (tangent) product $[x,y]\in T_e(G)$ is defined in component form by
\[
[x,y]^{i}\=c^{i}_{jk}x^{j}y^{k}=-[y,x]^{i},\qquad i=1,\dots,r.
\]
The tangent space $T_e(G)$ being equipped with such an anti-commutative
multiplication is called the \emph{tangent algebra} of the analytic
Moufang loop $G$. We shall use notation $\Gamma\=\{T_e(G),[\cdot,\cdot]\}$.

The tangent algebra of $G$ need not be a Lie algebra. There may exist such a triple
$x,y,z\in T_e(G)$ that does not satisfy the Jacobi identity:
\[
J(x,y,z)\=[x,[y,z]]+[y,[z,x]]+[z,[x,y]]\neq0.
\]
Instead, for any $x,y,z\in T_e(G)$ one has a more general \emph{Mal'tsev identity} \cite{Mal}
\[
[J(x,y,z),x]=J(x,y,[x,z]).
\]
Anti-commutative algebras with this identity are called the \emph{Mal'tsev algebras}.
Thus every Lie algebra is a Mal'tsev algebra as well.

\section{Birepresentations}

Consider a pair $(S,T)$ of the maps
$g\mapsto S_g$, $g\mapsto T_g$
of a Moufang loop $G$ into $GL_n$.
The pair $(S,T)$ is called a (linear) \emph{birepresentation}
of $G$  if the following conditions hold:
\begin{itemize}
\itemsep-2 pt
\item
$S_e=T_e=\1$,
\item
$T_gS_gS_h=S_{gh}T_g$,
\item
$S_gT_gT_h=T_{hg}S_g$.
\end{itemize}
The birepresentation $(S,T)$ is called \emph{associative}, if
the following simultaneous relations hold:
\[
S_gS_h=S_{gh},\quad T_gT_h=T_{hg},\quad S_gT_h=T_hS_g \qquad \forall\, g,h\in G
\]
In general, one can consider nonassociative birepresentations even for groups.

The \emph{generators} of a differentiable birepresentation $(S,T)$ of an analytic Moufang loop 
$G$ are defined as follows:
\[
S_j\=\frac{\p S_g}{\p g^{j}}\Bigg\vert_{g=e},
\quad
T_j\=\frac{\p T_g}{\p g^{j}}\Bigg\vert_{g=e},
\qquad
j=1,\dots,r.
\]

\begin{GLC}[\cite{Paal98,Paal04}]
The generators of a differentiable birepresentation of an analytic Moufang loop satisfy
the commutation relations
\begin{align*}
[S_j,S_k]&=2Y_{jk}+\frac{1}{3}c^p_{jk}S_p+\frac{2}{3}c^p_{jk}T_p\\
[S_j,T_k]&=-Y_{jk}+\frac{1}{3}c^p_{jk}S_p-\frac{1}{3}c^p_{jk}T_p\\
[T_j,T_k]&=2Y_{jk}-\frac{2}{3}c^p_{jk}S_p-\frac{1}{3}c^p_{jk}T_p
\end{align*}
where the Yamaguti operators $Y_{jk}$ obey the relations
\begin{gather*}
Y_{jk}+Y_{kj}=0\\
c^p_{jk}Y_{pl}+c^p_{jk}Y_{pl}+c^p_{jk}Y_{pl}=0.
\end{gather*}
and satisfy the reductivity relations
\begin{align*}
[Y_{jk},S_n]=d^p_{jkn}S_p\\
[Y_{jk},T_n]=d^p_{jkn}T_p
\end{align*}
and commutation relations
\[
[Y_{jk},Y_{ln}]=d^p_{jkl}Y_{pn}+d^p_{jkn}Y_{lp}
\]
with the Yamaguti constants
\[
6d^p_{jkl}\=c^p_{js}c^s_{kl}-c^p_{ks}c^s_{jl}+c^p_{sl}c^s_{jk}
\]
\end{GLC}

Dimension of this Lie algebra does not exceed
$2r+r(r-1)/2$. The Jacobi identities are guaranteed
by the defining identities of the Lie \cite{Loos} and general Lie \cite{Yam}
\emph{triple systems} associated with the tangent Mal'tsev algebra $T_e(G)$ of $G$.

These commutation relations are known from the theory of
alternative algebras \cite{Schafer}. This is due to the fact that commutator algebras of 
\emph{alternative} algebras are Mal'tsev algebras. 
In a sense, one can also say that the differential of a birepresentation  $(S,T)$ 
of the analytic Moufang loop is a \emph{birepresentation of its tangent Mal'tsev algebra}
$\Gamma$.

\section{Moufang-Noether currents and ETC}

Let us now introduce conventional canonical notations.
The coordinates of a space-time point $x$ are denoted by $x^{\al}$ ($\al=0,1,\dots,d-1$), where
$x^0=t$ is the time coordinate and $x^i$ are the spatial coordinates denoted concisely 
as $\x \= (x^1,\dots,x^{d-1})$. The Lagrange density $\L[u,\p u]$  is supposed to depend on a system of independent (bosonic or fermionic) fields $u^A(x)$ ($A=1,\ldots,n$) and their derivatives 
$\p_\al u^A\=u^A_\al$. The canonical $d$-momenta are denoted by
\[
p^\al_A\=\dfrac{\p\L}{\p u^A_\al}
\]
The Moufang-Noether currents are defined in vector (matrix) notations as follows:
\[
s^\al_j\=p^\al S_j u,\qquad t^\al_j\=p^\al T_j u
\]
and the corresponding  Moufang-Noether charges are defined as spatial integrals by
\[
\sigma_j(t)\=-\ii \int s^0_j(x) d\x, \qquad \tau_j(t)\=-\ii \int t^0_j(x) d\x
\]

By following canonical prescription assume that  the following equal-time commutators 
(or anti-commutators when $u^A$ are fermionic fields) hold:
\begin{align*}
[p^0_A(\x,t),u^B(\y,t)]&=-\ii \de_A^B \de(\x-\y)\\
[u^A(\x,t),u^B(\y,t)]&=0\\
[p^0_A(\x,t),p^0_B(\x,t)]&=0
\end{align*}
As a matter of fact, these equal-time commutators (ETC) do not depend on the associativity property of either $G$
nor $(S,T)$. Nonassociativity hides itself in the structure constants of $G$
and in the commutators $[S_j,T_k]$. Due to this, the computation of the ETC of the Noether-Moufang 
charge densities can be carried out in standard way and nonassociativity reveals itself only 
in the final step when the commutators $[S_j,T_k]$ are required.

First recall that in associative case the Noether charge densities obey the ETC
\begin{align*}
[s^0_j(\x,t),s^0_k(\y,t)]&=\hphantom{-}\ii c_{jk}^p  s^0_p(\x,t)\de(\x-\y)\\
[t^0_j(\x,t),t^0_k(\y,t)]&=-\ii c_{jk}^p t^0_p(\x,t)\de(\x-\y)\\
[s^0_j(\x,t),t^0_k(\y,t)]&=0
\end{align*}
It turns out that for non-associative Moufang transformations these ETC
are violated minimally. The Moufang-Noether charge density algebra reads
\begin{align*}
[s^0_j(\x,t),s^0_k(\y,t)]&=\hphantom{-}\ii c_{jk}^p s^0_p(\x,t)\de(\x-\y)
-2[s^0_j(\x,t),t^0_k(\y,t)]\\
[t^0_j(\x,t),t^0_k(\y,t)]&=-\ii c_{jk}^p t^0_p(\x,t)\de(\x-\y)
-2[s^0_j(\x,t),t^0_k(\y,t)]
\end{align*}
The ETCs 
\[
[s^0_j(\x,t),t^0_k(\y,t)]=[t^0_j(\y,t),s^0_k(\x,t)]
\]
represent associator of an analytic Moufang loop and so may be called the associator as well.
Associator of a Moufang loop is not arbitrary but have to fulfil certain constraints \cite{Paal98},
the \emph{generalized Lie and Maurer-Cartan equations}.
In the present sitiation the constraints can conveniently be listed  by
closing the above ETC, which in fact means construction of a \emph{finite} dimensional Lie
algebra generated by the Moufang-Noether charge densities.

Start by rewriting the Moufang-Noether algebra as follows:
\begin{align*}
[s^0_j(\x,t),s^0_k(\y,t)]
&=\ii\left[2Y^0_{jk}(x)+\frac{1}{3}c_{jk}^p s^0_p(x)+\frac{2}{3}c_{jk}^p t^0_p(x)\right]\de(\x-\y) \tag{1}\\
[s^0_j(\x,t),t^0_k(\y,t)]
&=\ii\left[-Y^0_{jk}(x)+\frac{1}{3}c_{jk}^p s^0_p(x)-\frac{1}{3}c_{jk}^p t^0_p(x)\right]\de(\x-\y)  \tag{2}\\
[t^0_j(\x,t),s^0_k(\y,t)]
&=\ii\left[2Y^0_{jk}(x)-\frac{2}{3}c_{jk}^p  s^0_p(x)-\frac{1}{3}c_{jk}^p t^0_p(x)\right]\de(\x-\y)  \tag{3}
\end{align*}
Here (2) can be seen as a definition of the \emph{Yamagutian} $Y^0_{jk}(x)$.
The Yamagutian is thus a recapitulation of the associator.
It can be shown that
\begin{gather*}
Y^0_{jk}(x)+Y^0_{kj}(x)=0  \tag{4}\\
c^p_{jk}Y^0_{pl}(x)+c^p_{kl}Y^0_{pj}(x)+c^p_{lj}Y^0_{pk}(x)=0 \tag{5}
\end{gather*}
The constraints (4) trivially descend from the anti-commutativity of the commutator
bracketing, but the proof of (5) needs certain effort. Further, it turns out
that the following \emph{reductivity} conditions hold:
\begin{align*}
[Y^0_{jk}(\x,t),s^0_n(\y,t)]=\ii d^p_{jkn}s^0_p(\x,t) \de(\x-\y) \tag{6}\\
[Y^0_{jk}(\x,t),s^0_n(\y,t)]=\ii d^p_{jkn}s^0_p(\x,t)  \de(\x-\y) \tag{7}
\end{align*}
Finally, by using the redictivity conditions, one can check that the 
Yamagutian obeys the Lie algebra
\begin{align*}
[Y^0_{jk}(\x,t),Y^0_{ln}(\y,t)]=\ii\left[d^p_{jkl}Y^0_{pn}(\x,t)+d^p_{jkn}Y^0_{lp}(\x,t)\right]\de(\x-\y) \tag{8}
\end{align*}
When integrating the above ETC (1) - (8) one can finally obtain the

\begin{thm}[Moufang-Noether charge algebra]
The Moufang-Noether charge algebra $(\sigma,\tau)$ is a birepresentation of the 
Mal'tsev algebra $\Gamma$.
\end{thm}

The paper was in part supported by the Estonian Science Foundation, Grant 5634.

\end{document}